\documentclass[]{article}

\usepackage{graphicx}
\usepackage{float}
\usepackage{url}
\usepackage{authblk}

\title{\textbf{SiD Simulation \& Analysis for ILC Snowmass Physics LoIs} \\ \vspace{0.2in} \small Talk presented at the International  Workshop on Future Linear Colliders (LCWS2021), 15-18 March 2021. C21-03-15.1}
\author[1,2]{Chris Potter\footnote{On behalf of the 2021 Snowmass SiD LoI authors:  Tim Barklow, Jim Brau, Lucas Braun, Masako Iwasaki, Laura Jeanty, Masakazu Kurata, Laura Nosler, Peter Onyisi, Austin Pryor, Amanda Steinhebel and Andy White}}
\affil[1]{Physics Department, University of Oregon}
\affil[2]{Institute for Fundamental Science, University of Oregon }

\date{\today}

\begin{document}

\maketitle

\begin{abstract}
The correct modeling of $e^+e^-$ collision events at the International Linear Collider (ILC), as well as the response of a collider detector like the Silicon Detector (SiD), is crucial to evaluating the expected sensitivity to key properties of the Higgs boson. In this document we describe the event generation and detector simulation in use for the SiD Letters of Interest submitted for the 2021 Snowmass community planning exercise.
\end{abstract}

\section{Introduction}

The 2012 discovery of the Higgs boson \cite{Aad:2012tfa, Chatrchyan:2012ufa} at the Large Hadron Collider (LHC) was a pivotal event in high energy particle physics. The community is moving toward a consensus that a Higgs factory, designed to enable the precision measurement of the Higgs properties, may be the best next step after the LHC. The International Linear Collider (ILC) \cite{Behnke:2013xla, Baer:2013cma, Phinney:2007gp, Behnke:2013lya} is one such Higgs factory. 

The Silicon Detector (SiD) is one of two detector concepts put forward in the ILC Technical Design Report (TDR) - the other is the International Large Detector (ILD) \cite{Behnke:2013lya}. SiD employs all Silicon tracking and electromagnetic calorimetry placed within a 5T solenoid to enable the particle flow technique \cite{Thomson200925}. While some of the sensitive detector technologies assumed for the TDR are being reconsidered in light of recent advances, nonetheless SiD provides a useful baseline design for evaluating the sensitivity to Higgs boson properties at the ILC Higgs factory.

The Snowmass community planning exercise \cite{snowmass2021}, sponsored by the Division of Particles and Fields of the American Physical Society, aims to bring physicists based in the US together with international partners in order to consider and evaluate potential next steps for their field. International partners have developed software tools for simulating Standard Model (SM) processes expected at the ILC assuming a generic ILC detector \cite{snowmass_ilc}. SiD is participating in this exercise.

In parallel, SiD is pursuing simulation tools specific to the SiD detector. In this document we describe the SiD simulation framework used for the signal and background processes targeted by the Snowmass Letters of Interest: Higgs to invisible \cite{loi_white}, Higgs to tau pairs \cite{loi_jeans}, Higgs to long-lived dark photons \cite{loi_jeanty}, and double Higgs production \cite{loi_potter}. Monte Carlo simulation files are available to Snowmass collaborators for download on the web upon approval by the SiD spokespeople \cite{sidmc20}.

\section{Common SiD Simulation Tools}

\subsection{Event Generation}

Two event generators are in use by SiD for generation of $e^+ e^-$ collision events: Whizard \cite{Kilian:2007gr} and MG5\_aMC@NLO \cite{Alwall:2014hca}. In addition to capturing the physics of the $e^+e^-$ interaction, Whizard also captures the crucial physics of initial state radiation (ISR) and beamstrahlung. Both effects involve radiation of photons from the initial state, lowering the effective center of mass energy and therefore impacting the cross section and event kinematics. At the time of writing MG5\_aMC@NLO has not yet implemented either ISR or beamstrahlung. 

However, both generators capture another critical feature of beams at the ILC: polarization. Polarization of the initial state electrons and positrons is central to the design of the ILC because it allows an additional constraint on the initial state and therefore also the final state. Moreover, initial states with righthanded electrons yield much smaller cross sections for some backgrounds. For all simulation, we assume the nominal TDR assumption of 80\% polarized electrons and 30\% polarized positrons. 

For final state showering and hadronization of events either Pythia6 \cite{Sjostrand:2006za} or Pythia8 \cite{Sjostrand:2007gs} are used. While Pythia6 is useful because it allows direct use of tunes obtained from the experiments at LEPII, Pythia8 is a far more flexible and modern tool despite the inapplicability of LEPII tunes.

\subsection{Detector Simulation}

\begin{figure}[t]
\begin{center}
\framebox{\includegraphics[width=1.575in, height=1.575in]{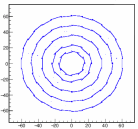}}
\framebox{\includegraphics[width=1.75in]{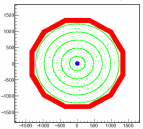}}
\framebox{\includegraphics[width=1.75in]{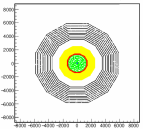}}
\end{center}
\caption{SiD barrel hitmaps produced in the $xy$ plane by Higgstrahlung events with full simulation in ILCSoft v02-00-02. Mapped are the Vertex Detector (left, blue hits), the Tracker and ECal (middle, green and red hits), and HCal and Muon detector (right, yellow and black hits). The units are millimeters (mm).}
\label{fig:sid}
\end{figure}

Fast detector simulation is useful for preliminary investigations of signal sensitivity because it is fast. It enables rapid identification of potential backgrounds and their order of magnitude after signal selection. To this end, we use the Delphes \cite{Mertens:2015kba} fast simulation of the SiD detector using the parametrized detector description in the DSiD card \cite{Potter:2016pgp}. DSiD directly implements the parametrized expected performance of SiD documented in the ILC TDR \cite{Behnke:2013xla}. 

Nevertheless, fast simulation has drawbacks which, in some cases, are severe. In these cases full detector simulation, for which all relevant physical processes are simulated in tracking individual particles through SiD, is critical. Even in cases where fast simulation reasonably capture full simulation performance, the latter should be used in any evaluation of sensitivity. For full SiD simulation we use the ILCSoft framework \cite{ilcsoft}, in which Geant4 \cite{AGOSTINELLI2003250} is employed using the DD4Hep interface \cite{frank_markus_2018_1464634}.

We use the compact XML description \texttt{SiD\_o2\_v03.xml} in the \texttt{lcgeo} package. See Figure \ref{fig:sid} for hitmaps describing the SiD barrel structure. We describe only the barrel here, but the endcap detector elements are similar. The Vertex Detector is implemented as a five barrel layers, radially from 13 mm to 59 mm, each layer described as sandwich of Carbon Fiber support with sensitive Silicon affixed to it with a layer of Epoxy. Containing the Vertex Detector, the Tracker is a also a five barrel layer structure, radially beginning at 250 mm, described as a sandwich of Carbon Fiber support, Epoxy, insensitive Silicon, sensitive Silicon, Kapton and Copper. 

The ECal is a twelve-sided polygon placed just outside the Tracker with 20 layers of thin (2.5 mm) and 10 layers of thick (5.0 mm) Tungsten absorber. These layers are described as a sandwich of Tungsten, Air, sensitive Silicon, Air, Copper, Kapton and Air. The HCal is also dodecahedral, with 11 layers of 19 mm Steel absorber and 3 mm sensitive plastic scintillator layers. The solenoid is placed outside of the calorimetry, and is described as Aluminum and Steel. Finally, the Muon detector is dodecahedral with Iron flux return in 196 mm layers, alternating with plastic scintillator described as a sandwich of Polysterene and reflecting Polystyrole foil.

\subsection{Object Reconstruction}

The Marlin (Modular Analysis and Reconstruction for the LINear Collider) reconstruction framework in ILCSoft-v02-00-02 is used for digitization of energy deposits in sensitive detector regions and reconstruction of tracks and calorimeter clusters. 

Tracks are reconstructed using \texttt{TruthTrackFinder} in the \texttt{MarlinTrkProcessors} package. The tracks are constructed and fitted from Vertex Detector and Tracker hits generated by digitization of energy deposits from charged particles made during the detector simulation stage. Thus the track fitting produces a genuine simulation of an experimental object as it is constructed and fitted from simulated hits. Hits are associated together using the Monte Carlo truth information, yielding trackfinding efficiency idealized yet reasonable given the performance achieved by trackfinding algorithms at the LHC.

The PandoraPFA \cite{Thomson200925} algorithm implements particle flow for SiD reconstruction and produces particle flow objects. The \texttt{DDPandoraPFANewProcessor} processor in the \texttt{DDMarlinPandora} package is used to cluster calorimeter hits and extrapolate tracks to the ECal, HCal and Muon detectors through the 5T SiD solenoid field. Here they are matched to clusters and identified as electrons, charged hadrons, or muons (respectively) provided the matching criteria are met. Jet energy resolution is thus improved throught the use of the more precise tracking momentum measurement rather than the calorimeter energy measurement. Clusters in the ECal and HCal which are left unmatched to tracks are assumed to be photons or neutral hadrons (respectively).

\section{Common Background Simulation}

The challenge of generating a full set of ILC SM process simulation is described in \cite{Berggren:2021sju}. The set generated for the Detailed Baseline Design (DBD) common to SiD and ILD was used for studies documented in the ILC TDR \cite{Behnke:2013lya}. This set of Whizard 1.40 files in StdHep \cite{stdhep} format has been preserved and is in use for Snowmass 2021. These include files for four center of mass energies ($\sqrt{s}=250,350,500,1000$~GeV) and include all processes above threshold with final state fermion multiplicities from one to eight, assuming 100\% polarized beams $e_L^- e_R^+$ and $e_R^-e_L^+$. This includes processes initiated from an $e\gamma$ initial state, which yields odd final state fermion multiplicities.

\subsection{Barklow DBD \texttt{all\_SM\_background}}

Because many distinct processes were generated for the DBD, each in separate files, and with pure polarized beams, they must be mixed together in the correct way to simulate the SM and the assumed beam polarization fractions. During the DBD study this was done once for SiD at the StdHep level by Tim Barklow in order to streamline analysis for end users. Each process was weighted by SM cross section and polarization weights necessary to reproduce 80\% polarized electron and 30\% polarized positron beams. These mixed samples are named \texttt{all\_SM\_background}.

Thus detector simulation, whether fast or full, and with any software and detector version, can simply run on these StdHep files to simulate the full SM at the ILC assuming the nominal 80/30 polarization scheme. Each file contains all processes, though in order to alleviate storage issues and to speed simulation and analysis events are weighted with weights larger than one. In some cases, SM processes which are interesting signals (\emph{e.g.} Higgs or top pair production) have been separated out from the mixed samples, in which case the mixed samples are referred to as \texttt{all\_other\_SM\_background}.

See Table \ref{tab:allsm} for some details on the Barklow mixed DBD background files. All of the files in this table have been passed through the Delphes 3.4.2 fast simulation with the DSiD card and full simulation of SiD option 2 version 3, in ILCSoft v02-00-02.

\begin{table}
\begin{center}
\begin{tabular}{|c|c|c|c|c|} \hline
$\sqrt{s}$ [GeV] & Filename & $\mathcal{L}$ [fb$^{-1}$] & Final States & Signals\\ \hline
250 & \texttt{all\_SM\_background} & 250 & 1f,2f,3f,4f & \texttt{higgs\_ffh} \\ 
350 & \texttt{all\_other\_SM\_background} & 350 & 1f,2f,3f,4f & \texttt{higgs\_ffh},\texttt{ttbar} \\ 
500 & \texttt{all\_SM\_background} & 500 & 1f,2f,3f,4f & \texttt{6f\_ttbar\_mt173p5} \\ 
1000 & \texttt{all\_other\_SM\_background} & 1000 & 1f,2f,3f,4f,5f,6f,7f,8f & [many] \\ \hline
\end{tabular}
\caption{The Barklow mixed DBD SM background Whizard samples and their assumed center of mass energy, integrated luminosity and final state fermion multiplicities. These have been passed through the Delphes fast simulation v3.4.2 with the DSiD card and full simulation of SiD option 2 version 3, in ILCSoft v02-00-02}
\label{tab:allsm}
\end{center}
\end{table}

\subsection{SiD MC20 and MC21 Campaigns}

While the mixed DBD background files are useful for identifying background processes in cases of large background yields, they are statistically limited and in some cases the process event weight is prohibitively large. For such cases new background samples have been produced in the SiD MC20 and MC21 production campaigns using Whizard 2.6.4 with the polarization fractions set to the nominal TDR design assumption. The equivalent luminosity is $\int dt \mathcal{L}=2$~ab$^{-1}$. In these samples ISR is simulated but beamstrahlung is not. 

The effect of beamstrahlung is important and must be considered in any sensitivity estimate. Beamstrahlung has been neglected in these samples because the effect depends critically on the beam parameters, which will not be known at least until the ILC design has been finalized. However, Whizard can include the effect if the beam parameters are known. For examples of the impact of beamstrahlung on background processes see Table 6 in \cite{Potter2020PrimerOI}. Systematic studies to evaluate the impact of beamstrahlung will be performed for the Snowmass studies discussed here.

At the time of writing, full SM samples for 2f,3f,4f (,6f) final states have been produced for $\sqrt{s}=250$~GeV ($\sqrt{s}=500$~GeV). The MC21 production is ongoing and new samples will be produced in due course. These MC20 and MC21 files have been passed through the Delphes 3.4.2 fast simulation with the DSiD card and full simulation of SiD option 2 version 3, in ILCSoft v02-00-02.

\section{Signal Simulation and Analysis}

\begin{figure}[t]
\begin{center}
\includegraphics[width=\textwidth]{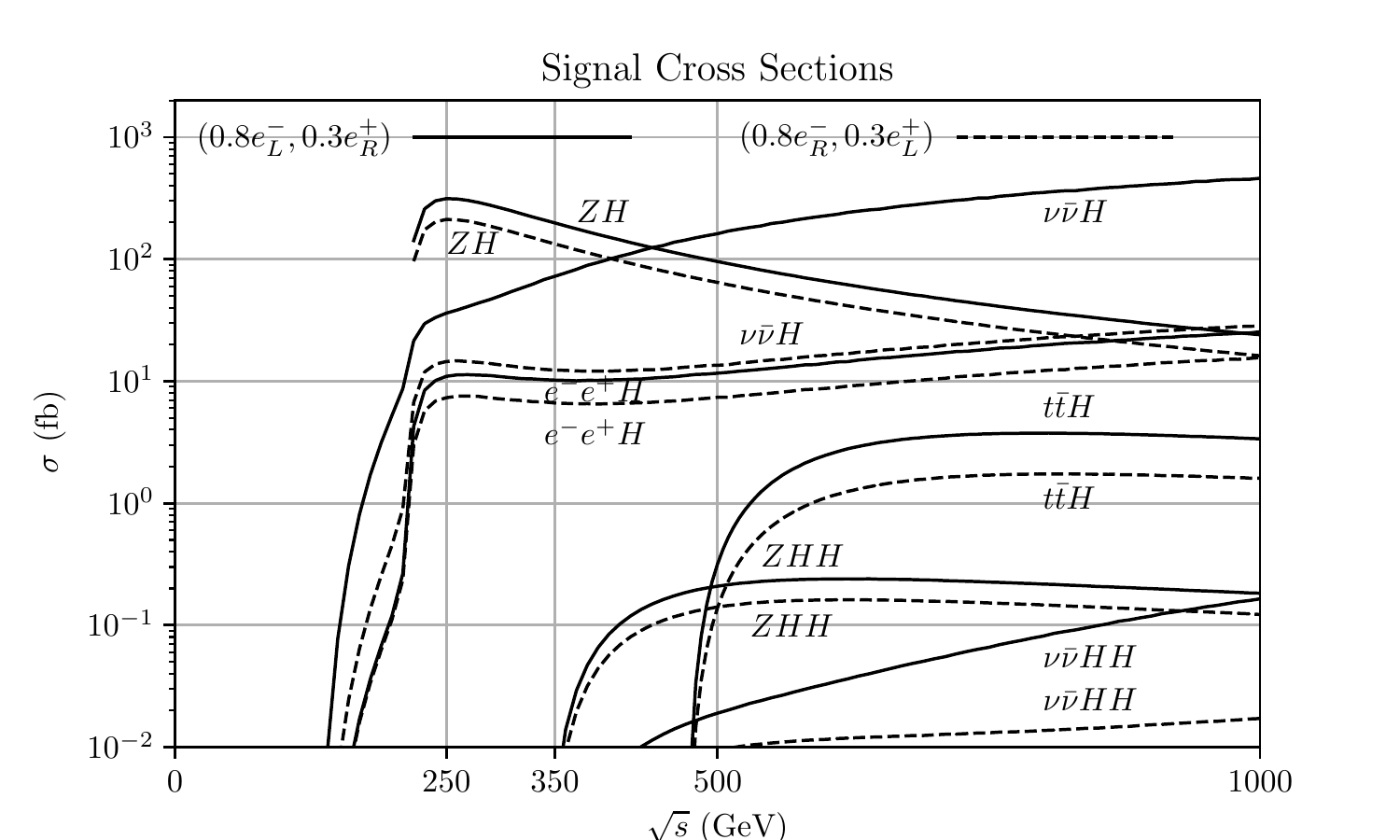}
\caption{Cross sections for signal Higgs production processes vs. $\sqrt{s}$, assuming the nominal ILC TDR beam polarization, obtained with Whizard 2.6.4 \cite{Kilian:2007gr}. ISR is included but beamstrahlung is not. Taken from \cite{Potter2020PrimerOI}.}
\label{fig:hxsec}
\end{center}
\end{figure}

Higgs bosons are produced in $e^+e^-$ collisions in several ways. Higgstrahlung $e^+e^- \rightarrow ZH$ is an $s$-channel process which peaks near $\sqrt{s}=250$~GeV. The $t$-channel vector boson fusion processes $e^+e^- \rightarrow \nu \bar{\nu}H, e^+e^-H$ supplement Higgstrahlung at higher $\sqrt{s}$. Higgs bosons can also be produced in association with a top quark pair, $e^+e^- \rightarrow t\bar{t}H$ above a threshold near $\sqrt{s}=500$~GeV. In double Higgs production processes $e^+e^- \rightarrow ZHH,\nu\bar{\nu}HH$ two Higgs bosons are produced in association either with a $Z$ boson or a neutrino pair. See Figure \ref{fig:hxsec} for signal cross sections vs. $\sqrt{s}$.

For all SiD Snowmass LoIs except the Higgs self-coupling LoI, the Higgstrahlung process is chosen at $\sqrt{s}=250$~GeV. The yields at this $\sqrt{s}$ ensure high precision measurements, though these analyses could be performed at any $\sqrt{s}$ above thresholds for Higgs production.  In all cases the event generation files have been passed through the Delphes fast simulation v3.4.2 with the DSiD card and full simulation of SiD option 2 version 3, in ILCSoft v02-00-02.

\subsection{Higgs to Invisible}

In the SM the Higgs boson decays invisibly when it decays through $H \rightarrow ZZ^{\star} \rightarrow \nu \bar{\nu} \nu \bar{\nu}$ since neutrinos interact only weakly and cannot be detected in a collider detector. In theories of Beyond the SM (BSM) physics, the Higgs may decay to other particles which are not detected in a collider detector, \emph{e.g.} dark matter. 

The SiD MC20 event generation for the analysis to evaluate the expected precision for invisible Higgs decays with SiD is performed with Whizard 2.6.4. Higgstrahlung $e^+e^- \rightarrow ZH$ at $\sqrt{s}=250$~GeV is assumed with the nominal beam polarization, and the Higgs is required to decay in the SM channel to four neutrinos. Thus this sample may be used to study sensitivity for any model in which the Higgs decays to BSM particles which are not detected by a collider detector. For a description of the analyses of the invisible Higgs analyses in progress for Snowmass, see \cite{steinhebel2021hrightarrowinvisible}.

\subsection{Higgs to Long-lived Dark Photons}

In the SM the Higgs boson may decay to long-lived particles through quark pairs $H \rightarrow q\bar{q}$ when the quarks hadronize into mesons or baryons which do not decay promptly, \emph{e.g.} $B$, $D$ and $K$ mesons. In a BSM model like the dark photon model \cite{Curtin_2015}, the Higgs can decay through $H \rightarrow Z_D Z_D, Z Z_D, H_D H_D$, where $Z_D$ are dark photons which then decay to fermion pairs and $H_D$ is the dark singlet Higgs. 

In the dark photon model there are four free parameters: the $Z_D$ mass and $\gamma/Z_D$ mixing parameter $\epsilon$ and the $H_D$ mass and $H/H_D$ mixing parameter $\kappa$. The lifetime of the $Z_D$ depends on $m_{Z_D}$ and $\epsilon$ and for suitable ranges of these parameters can produce non-prompt $Z_D$ decays. The SiD MC20 event generation for the analysis to evaluate the expected sensitivity to such decays is performed with MG5\_aMC@NLO using the HAHM model \cite{hahm} written by the authors of \cite{Curtin_2015}.

It should be emphasized that while the dark photon model is a very specific model, the phenomenology of \emph{any} long-lived particle decaying to fermion pairs (\emph{e.g.} axions) in a collider detector is captured by the events produced with that model.

\subsection{Higgs CP in Higgs to Tau Pairs}

This Snowmass LoI includes authors from both ILD and SiD. For the SiD authors samples have been generated assuming the SM. The event generation is performed with MG5\_aMC@NLO with Pythia8 for hadronization and tau decay. BSM values for the CP angles can be set in MG5\_aMC@NLO using alternate models such as the Minimal Supersymmetric SM (MSSM) or Next to MSSM (NMSSM), which adds a Higgs singlet to the five Higgs bosons of the MSSM. SiD MC20 and MC21 production is ongoing.

The proper treatment of tau decays is critical to the correct simulation of $H \rightarrow \tau^+ \tau^-$ events. Spin correlations must be preserved from the tau to daughter particles and throughout the decay chain. The treatment of tau decay in Pythia8 has been validated against other tools like Tauola, so Pythia8 is invoked for tau decay as well as hadronization.

\subsection{Higgs Self-coupling}

Since the Higgs self-coupling cannot be accessed in single Higgs production modes, the double Higgs productions modes $e^+e^- \rightarrow ZHH$ at $\sqrt{s}=500$ and $e^+e^- \rightarrow \nu \bar{\nu}HH$ at $\sqrt{s}=1000$~GeV are assumed in the event generation. 

The self-coupling is assumed to be the SM coupling, but any BSM coupling will affect double Higgs events primarily throught the impact on the cross section. Therefore the SM samples can be used for BSM studies if the BSM model does not modify the coupling in such a way as to modify event kinematics. The Higgs decays are fully inclusive assuming SM branching ratios. Whizard 2.6.4 is used for the LoI study.

\section{Conclusion}

We have described the Monte Carlo simulations prepared by the SiD consortium for Letters of Interest submitted to the Snowmass 2021 community planning exercise. These are collider events with $e^+e^-$ and $e\gamma$ initial states at $\sqrt{s}=250,350,500,1000$~GeV generated by Whizard and MG5\_aMC@NLO. The SiD detector response is simulated with Delphes fast simulation and ILCSoft full simulation. Samples generated for the DBD exercise with the full SM for background studies are available, together with new dedicated samples for high statistics background studies. Signal processes generated and simulated for Snowmass LoIs include Higgstrahlung with Higgs boson decays to invisible final states, long-lived particles and tau pairs. Double Higgs production processes are also included, with inclusive SM Higgs decays.

\bibliography{paper}

\begin{thebibliography}{10}

\bibitem{Aad:2012tfa}
Georges Aad et~al.
\newblock {Observation of a new particle in the search for the Standard Model
  Higgs boson with the ATLAS detector at the LHC}.
\newblock {\em Phys.Lett.}, B716:1--29, 2012, arXiv:1207.7214.

\bibitem{Chatrchyan:2012ufa}
Serguei Chatrchyan et~al.
\newblock {Observation of a new boson at a mass of 125 GeV with the CMS
  experiment at the LHC}.
\newblock {\em Phys.Lett.}, B716:30--61, 2012, arXiv:1207.7235.

\bibitem{Behnke:2013xla}
Ties Behnke, James~E. Brau, Brian Foster, Juan Fuster, Mike Harrison,
  James~McEwan Paterson, Michael Peskin, Marcel Stanitzki, Nicholas Walker, and
  Hitoshi Yamamoto.
\newblock {The International Linear Collider Technical Design Report - Volume
  1: Executive Summary}.
\newblock 2013, arXiv:1306.6327.

\bibitem{Baer:2013cma}
Howard Baer, Tim Barklow, Keisuke Fujii, Yuanning Gao, Andre Hoang, Shinya
  Kanemura, Jenny List, Heather~E. Logan, Andrei Nomerotski, and Maxim
  Perelstein.
\newblock {The International Linear Collider Technical Design Report - Volume
  2: Physics}.
\newblock 2013, arXiv:1306.6352.

\bibitem{Phinney:2007gp}
Gerald Aarons et~al.
\newblock {ILC Reference Design Report Volume 3 - Accelerator}.
\newblock 2007, arXiv:0712.2361.

\bibitem{Behnke:2013lya}
Ties Behnke, James~E. Brau, Philip~N. Burrows, Juan Fuster, Michael Peskin,
  Marcel Stanitzki, Yasuhiro Sugimoto, Sakue Yamada, and Hitoshi Yamamoto.
\newblock {The International Linear Collider Technical Design Report - Volume
  4: Detectors}.
\newblock 2013, arXiv:1306.6329.

\bibitem{Thomson200925}
M.A. Thomson.
\newblock {Particle flow calorimetry and the PandoraPFA algorithm}.
\newblock {\em Nuclear Instruments and Methods in Physics Research Section A:
  Accelerators, Spectrometers, Detectors and Associated Equipment}, 611(1):25
  -- 40, 2009.

\bibitem{snowmass2021}
APS DPF.
\newblock {Snowmass 2021}.
\newblock \url{https://snowmass21.org/}.
\newblock Accessed: May, 2021.

\bibitem{snowmass_ilc}
{ILC Simulation Resources for Snowmass 2021}.
\newblock \url{http://ilcsnowmass.org/}.
\newblock Accessed: May, 2021.

\bibitem{loi_white}
Andrew White, Austin Prior, James Brau, Christopher Potter, Amanda Steinhebel,
  and Makayla Massar.
\newblock {LOI - ILC/SiD Higgs to Invisible}.
\newblock
  \url{https://www.snowmass21.org/docs/files/summaries/EF/SNOWMASS21-EF2_EF1_Andy_White\%2C_Jim_Brau-185.pdf}.
\newblock Accessed: May, 2021.

\bibitem{loi_jeans}
D.~Jeans, I.~Bozovic-Jelisavcic, G.~Milutinovic-Dumbelovic, J.~Brau, L.~Braun,
  and C.~Potter.
\newblock {Measuring the CP Properties of the Higgs Sector at Electron-Positron
  Colliders}.
\newblock
  \url{https://www.snowmass21.org/docs/files/summaries/EF/SNOWMASS21-EF1_EF2_DanielJeans-113.pdf}.
\newblock Accessed: May, 2021.

\bibitem{loi_jeanty}
Laura Jeanty, Laura Nosler, and Chris Potter.
\newblock {Sensitivity to Long-lived Dark Photons at the ILC}.
\newblock
  \url{https://www.snowmass21.org/docs/files/summaries/EF/SNOWMASS21-EF9_EF8-081.pdf}.
\newblock Accessed: May, 2021.

\bibitem{loi_potter}
Tim Barklow, James Brau, Masako Iwasaki, Masakazu Kurata, Peter Onyisi, and
  Chris Potter.
\newblock {Higgs Self-coupling at the ILC with the SiD Detector}.
\newblock
  \url{https://www.snowmass21.org/docs/files/summaries/EF/SNOWMASS21-EF1_EF2_Potter-155.pdf}.
\newblock Accessed: May, 2021.

\bibitem{sidmc20}
{SiD Monte Carlo Exercise 2020/2021}.
\newblock \url{https://pages.uoregon.edu/ctp/SiD_private.html}.
\newblock Accessed: May, 2021.

\bibitem{Kilian:2007gr}
Wolfgang Kilian, Thorsten Ohl, and Jurgen Reuter.
\newblock {WHIZARD: Simulating Multi-Particle Processes at LHC and ILC}.
\newblock {\em Eur. Phys. J.}, C71:1742, 2011, arXiv:0708.4233.

\bibitem{Alwall:2014hca}
J.~Alwall, R.~Frederix, S.~Frixione, V.~Hirschi, F.~Maltoni, O.~Mattelaer,
  H.~S. Shao, T.~Stelzer, P.~Torrielli, and M.~Zaro.
\newblock {The automated computation of tree-level and next-to-leading order
  differential cross sections, and their matching to parton shower
  simulations}.
\newblock {\em JHEP}, 07:079, 2014, arXiv:1405.0301.

\bibitem{Sjostrand:2006za}
Torbjorn Sjostrand, Stephen Mrenna, and Peter~Z. Skands.
\newblock {PYTHIA 6.4 Physics and Manual}.
\newblock {\em JHEP}, 0605:026, 2006, hep-ph/0603175.

\bibitem{Sjostrand:2007gs}
Torbjorn Sjostrand, Stephen Mrenna, and Peter~Z. Skands.
\newblock {A Brief Introduction to PYTHIA 8.1}.
\newblock {\em Comput.Phys.Commun.}, 178:852--867, 2008, arXiv:0710.3820.

\bibitem{Mertens:2015kba}
Alexandre Mertens.
\newblock {New features in Delphes 3}.
\newblock {\em J. Phys. Conf. Ser.}, 608(1):012045, 2015.

\bibitem{Potter:2016pgp}
C.~T. Potter.
\newblock {DSiD: a Delphes Detector for ILC Physics Studies}.
\newblock In {\em {Proceedings, International Workshop on Future Linear
  Colliders (LCWS15): Whistler, B.C., Canada, November 02-06, 2015}}, 2016,
  arXiv:1602.07748.

\bibitem{ilcsoft}
GitHub.
\newblock {ILCSoft}.
\newblock \url{https://github.com/iLCSoft}.
\newblock Accessed: August 10, 2021.

\bibitem{AGOSTINELLI2003250}
S.~Agostinelli et~al.
\newblock {Geant4 — a simulation toolkit}.
\newblock {\em Nuclear Instruments and Methods in Physics Research Section A:
  Accelerators, Spectrometers, Detectors and Associated Equipment}, 506(3):250
  -- 303, 2003.

\bibitem{frank_markus_2018_1464634}
Markus Frank, Frank Gaede, Marko Petric, and Andre Sailer.
\newblock {AIDASoft/DD4hep}, 2018.
\newblock \url{http://dd4hep.cern.ch/}.

\bibitem{Berggren:2021sju}
Mikael Berggren.
\newblock {Generating the full SM at linear colliders}.
\newblock {\em PoS}, ICHEP2020:903, 2021, arXiv:2105.04049.

\bibitem{stdhep}
L.~Garren.
\newblock {StdHep 5.06.01 Monte Carlo Standardization at FNAL Fortran and C
  Implementation}.
\newblock
  \url{http://cd-docdb.fnal.gov/0009/000903/015/stdhep_50601_manual.ps}.
\newblock Accessed: March 14, 2016.

\bibitem{Potter2020PrimerOI}
C.~Potter.
\newblock {Primer on ILC physics and SiD software tools}.
\newblock {\em The European Physical Journal Plus}, 135:1--63, 2020,
  arXiv:2002.02399.

\bibitem{steinhebel2021hrightarrowinvisible}
Amanda Steinhebel, Jim Brau, and Chris Potter.
\newblock H$\rightarrow$invisible at the ilc with sid, 2021, arXiv:2105.00128.

\bibitem{Curtin_2015}
David Curtin, Rouven Essig, Stefania Gori, and Jessie Shelton.
\newblock Illuminating dark photons with high-energy colliders.
\newblock {\em Journal of High Energy Physics}, 2015(2), Feb 2015.

\bibitem{hahm}
{SM + Dark Vector + Dark Higgs Madgraph5 Model}.
\newblock \url{http://insti.physics.sunysb.edu/~curtin/hahm_mg.html}.
\newblock Accessed: May, 2021.

\end{thebibliography}

\end{document}